\documentclass[reqno,11pt]{amsart}
\usepackage[
  a4paper,
  top=26mm,
  bottom=26mm,
  left=26mm,
  right=26mm
]{geometry}

\usepackage{graphicx}

\usepackage{amsmath,amssymb,amscd}
\usepackage[mathscr]{eucal}
\usepackage{here}
\usepackage{bm}

\usepackage{tikz}
\usetikzlibrary{arrows.meta,calc,decorations.markings,math,positioning}

\newtheorem{thm}{Theorem}[section]
\newtheorem{cor}[thm]{Corollary}
\newtheorem{lem}[thm]{Lemma}
\newtheorem{prop}[thm]{Proposition}

\theoremstyle{definition}

\theoremstyle{remark}
\newtheorem{remark}[thm]{Remark}

\def\C{{\mathbb C}}
\def\Z{{\mathbb Z}}
\def\ve{{\varepsilon}}

\newcommand\LL{{\mathcal{L}}}

\newcommand{\uu}{{\mathsf u}}
\newcommand{\ww}{{\mathsf w}}

\newcommand{\gf}{{\mathcal G}}
\newcommand{\st}{{\mathcal S}}
\newcommand{\wt}{{\mathsf{wt}}}
\newcommand{\dd}{{\mathsf d}}
\newcommand{\cc}{{\mathcal C}}

\newcommand{\hf}{\mathsf{h}}

\newcommand{\maru}[1]{\raise0.2ex\hbox{\textcircled{\scriptsize{#1}}}}

\newfont{\bg}{cmr9 scaled\magstep4}
\newcommand{\bigzerol}{\smash{\lower1.0ex\hbox{\bg 0}}}

\tikzset{
  mid arrow/.style={postaction={decorate,decoration={
        markings,
        mark=at position .5 with {\arrow[#1]{latex}}
      }}},
}

\begin{document}

\title[Quantized six-vertex model]
{Quantized six-vertex realizations of frustration graphs}

\dedicatory{Dedicated to the memory of Rodney James Baxter}

\author[Atsuo Kuniba]{Atsuo Kuniba}
\address{Atsuo Kuniba, Graduate School of Arts and Sciences, 
University of Tokyo, Komaba, Tokyo 153-8902, Japan.}
\email{atsuo.s.kuniba@gmail.com}

\author[Souta Oikawa]{Souta Oikawa}
\address{Souta Oikawa, Graduate School of Arts and Sciences, 
University of Tokyo, Komaba, Tokyo 153-8902, Japan.}
\email{4639042647@g.ecc.u-tokyo.ac.jp}

\begin{abstract}
A recent graph-theoretic criterion for a Hamiltonian to exhibit a
free-parafermion spectrum is that its frustration graph be an
indifference graph.
We show that an arbitrary finite indifference graph can be realized
as a frustration graph of the quantized six-vertex model at odd
roots of unity, by a suitable choice of the defining wiring diagram
and parameters.
This provides a concrete realization of the criterion through the
quantum Frobenius property of the quantized six-vertex model, 
recently established from its three-dimensional integrable structure.
\end{abstract}

\maketitle

\section{Introduction}

In his seminal work \cite{B89}, Rodney Baxter discovered an exactly
solvable $\Z_N$-Hamiltonian whose spectrum takes the form
\begin{equation}
\omega^{n_1}\mathcal E_1+\cdots+\omega^{n_d}\mathcal E_d,
\end{equation}
where $\omega$ is a primitive $N$th root of unity and $n_i\in\Z_N$.
For general $N$, the spectrum is non-Hermitian and provides a natural
$\Z_N$-generalization of the familiar Ising-type free-fermion spectrum
$\pm\mathcal E_1\pm\cdots\pm\mathcal E_d$, recovered at $N=2$.
The original derivation proceeded through an indirect connection with the
superintegrable chiral Potts model \cite{B89,B89b}, and
the mechanism underlying this remarkably simple spectrum remained 
mysterious for many years.

A breakthrough came with \cite{F14}, where the relevant integrability 
was investigated under the name of \emph{free parafermions}.
Subsequent works \cite{B14,AYP14,AYP16} developed similar algebraic
approaches for the more general $\tau_2$ model \cite{BS90,BBP90,B04}, 
again closely related to the
chiral Potts model and containing the free-parafermion model as a
special case.

These developments, still technically rather demanding, naturally raise a question:
can free-parafermion spectra arise in a broader class of models?
This was answered affirmatively in \cite{F19,ECF21,AP2,A25};
see also the concise review \cite{BHL23}.
For a Hamiltonian written as a sum of local interaction terms,
\begin{equation}
H=h_1+\cdots+h_M,
\end{equation}
the free-parafermion spectrum can follow directly from suitable
$\omega$-commutation relations among the operators
$h_1,\ldots,h_M$, without reference to the chiral Potts construction.
Such relations can be realized in one-dimensional spin chains with
multispin interactions of increasingly large range.

This perspective was recently systematized in a
graph-theoretic framework in \cite{MEWC25}.
There, the local terms $h_i$ are associated with an oriented graph $\gf$,
called the \emph{frustration graph}, which records the
$\omega$-commuting pairs; see \eqref{halg}.
A central object in this approach is the \emph{independent-set operator},
defined by
\begin{equation}
\mathcal T_{\gf}(y)
=
\sum_{I\in\mathcal I(\gf)}
y^{|I|}
\prod_{i\in I}h_i,
\end{equation}
where $\mathcal I(\gf)$ denotes the collection of independent sets of
$\gf$.
Here we use the variable $y$, which corresponds to $u=-y$ in the
convention of \cite{MEWC25}.
When $\gf$ is an \emph{indifference graph}
and each $h_i^N$ is proportional to the identity,
the operators $\mathcal T_{\gf}(y)$ form a commuting family and the
free-parafermion spectrum of the Hamiltonian follows.
Thus the result gives a graph-theoretic criterion in terms of
the frustration graph and the local operator relations, which is
independent of any particular spin-chain realization.

Around the same time, a new connection between free-parafermion models
and the quantized six-vertex (Q6V) model was beginning to emerge.
The Q6V model was introduced in \cite{IKTY25} on \emph{wiring diagrams} $G$,
based on the $q$-Weyl algebra and three-dimensional integrability.
At a primitive odd $N$th root of unity $q=\varepsilon$, its commuting
layer transfer matrix was subsequently shown to satisfy the functional
relation \cite{IKTY26}
\begin{equation}
\mathbb T_G(y)\mathbb T_G(\varepsilon y)\cdots
\mathbb T_G(\varepsilon^{N-1}y)
=
\mathscr P_{G,N}(y^N)\mathrm{Id},
\end{equation}
where $\mathscr P_{G,N}$ is the corresponding scalar polynomial.
Generalized inversion relations of this kind had been recognized in
earlier works as a key mechanism underlying free-parafermion spectra.
The result of \cite{IKTY26} shows that this functional relation is a
manifestation of a \emph{quantum Frobenius property}, in the spirit of
the quantum Frobenius phenomenon for quantum groups at roots of unity \cite{L90}.  
It is an inherent property shared by a broad class of genuinely
two-dimensional Q6V models, yet its origin can be traced to
three-dimensional integrability, most notably to the
\emph{tetrahedron equations}.

The aim of the present paper is to realize the graph-theoretic framework
for free-parafermion spectra from the Q6V model described above.
Our main result is that, by a suitable choice of the wiring diagram $G$
and of the parameters of the Q6V model,
an arbitrary finite indifference graph $\gf$ can be realized
as its frustration graph.
Moreover, the layer transfer matrix $\mathbb T_G(y)$
yields the independent-set operator $\mathcal T_\gf(y)$.
Thus the Q6V construction covers the full class of
indifference-graph systems considered in \cite{MEWC25}.

Our strategy is to focus on the particular family of
wiring diagrams $G_{p,L}$ shown in Figure~\ref{f:grid}, which turns out
to be sufficient for our purpose.
For generic parameters, the elementary path weights,
which realize the operators $h_1,\ldots,h_M$ above, have the \emph{path power}
$P_M^p$ as their frustration graph.
A useful graph-theoretic ingredient here is that every finite
indifference graph $\gf$ can be realized, with its prescribed ordering,
as an induced subgraph of a suitable path power.
By setting appropriate Q6V parameters to zero, one can therefore
select precisely the vertices corresponding to $\gf$.

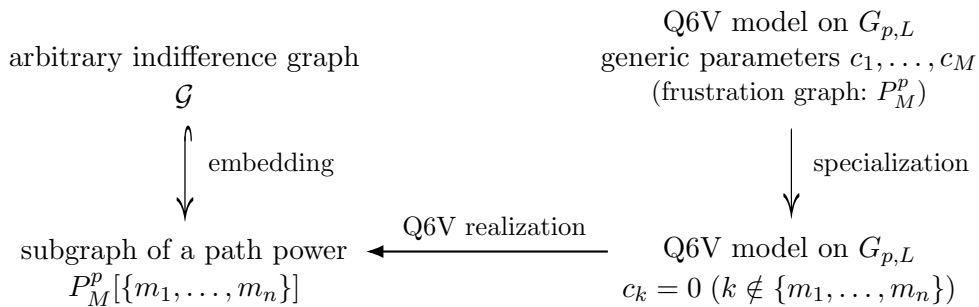
\begin{figure}[H]
\centering
\begin{tikzpicture}[x=1cm,y=1cm,>=Latex]

\node (G1) at (1.2,3.00) {arbitrary indifference graph};
\node (G2) at (1.2,2.50) {$\gf$};

\node (PMtitle) at (1.2,0.45) {subgraph of a path power};
\node (PM) at (1.2,-0.05) {$P_M^p[\{m_1,\ldots,m_n\}]$};

\node (Qgen) at (9.2,3.0) {%
\begin{tabular}{c}
Q6V model on $G_{p,L}$\\
generic parameters $c_1,\ldots,c_M$\\ 
{\small (frustration graph:~$P_M^p$)}
\end{tabular}
};

\node (Qspec1) at (9.2,0.45) {Q6V model on $G_{p,L}$};
\node (Qspec2) at (9.2,-0.05) {$c_k=0\;(k\notin\{m_1,\ldots,m_n\})$};


\node at (1.2,1.48)
  {\rotatebox[origin=c]{-90}{\scalebox{3.3}[1.1]{$\hookrightarrow$}}};
\node at (2.35,1.6) {\small embedding};

\node at (9.2,1.48)
  {\rotatebox[origin=c]{-90}{\scalebox{3.5}[1.05]{$\rightarrow$}}};
\node at (10.55,1.6) {\small specialization};

\draw[->,line width=0.8pt] (6.8,0.45) -- (3.6,0.45);
\node at (5.3,0.75) {\small Q6V realization};

\end{tikzpicture}
\caption{Schematic outline of the Q6V realization of an arbitrary finite
indifference graph through its frustration graph, where $M$ is given by \eqref{M}.
The graph-theoretic ingredients in the left column are treated in
Section~\ref{s:graph}, while the Q6V construction in the right column is discussed
in Sections~\ref{s:Q6V} and~\ref{s:main}.}
\label{f:str}
\end{figure}

With this choice of parameters, there is a degree- and weight-preserving
bijection between the independent sets $\mathcal I(\gf)$ and the Q6V states
of nonzero weight.
Consequently, the independent-set operator
$\mathcal T_{\gf}(y)$ is identified with the corresponding Q6V
transfer matrix.
At roots of unity, this identification allows the commutativity and
Frobenius relations for $\mathcal T_{\gf}(y)$ to be recovered directly
from the corresponding Q6V identities.
The overall scheme is summarized in Figure~\ref{f:str}.

The paper is organized as follows.
Section~\ref{s:graph} reviews and develops the graph-theoretic ingredients
used below, including indifference graphs, path powers, and independent-set
operators.
Section~\ref{s:Q6V} introduces the Q6V model on the wiring diagrams
$G_{p,L}$ and its elementary path weights.
It then establishes a degree- and weight-preserving correspondence
between Q6V states of nonzero weight and independent sets of induced
subgraphs of path powers.
Section~\ref{s:main} specializes the construction to roots of unity and
proves the Q6V realization of arbitrary finite indifference graphs and
their independent-set operators.
Section~\ref{s:o} summarizes the results and discusses the relation
between the graph-theoretic and Q6V frameworks, together with their
respective ranges of generality and some open problems.
Appendix~\ref{app:embed} gives an explicit construction of the
path-power embedding of a finite indifference graph, while
Appendix~\ref{app:nhr} discusses non-homogeneous range
representations and their equivalence with finite indifference graphs.

\section{Indifference graphs and independent-set operators}\label{s:graph}

This section introduces the graph-theoretic structures that will be
used in the Q6V realization developed below.
We review indifference graphs and path powers, including the realization
of a finite indifference graph as an induced subgraph of a path power.
A related description in terms of non-homogeneous ranges is given in
Appendix~\ref{app:nhr}.
We then recall, following \cite{MEWC25}, the independent-set operators
associated with an oriented indifference graph and their basic properties.

\subsection{Basic definitions}\label{ss:bd}

We recall the standard graph-theoretic notions used throughout the paper;
see, for example,~\cite{D25}.
All graphs considered in this paper are finite and simple, that is,
they have neither loops nor multiple edges.

For a graph $\gf=(V,E)$, we let $V(\gf)=V$ and $E(\gf)=E$ denote the sets of
vertices and edges.
Unless otherwise stated, graphs are understood to be undirected, so
an edge is an unordered pair $\{u,v\}$ of distinct vertices.
Two vertices $u,v\in V$ are said to be \emph{adjacent} if
$\{u,v\}\in E$.

We shall later orient the edges of an indifference graph according to
a chosen indifference ordering.
Until then, all graph-theoretic notions refer to the underlying
undirected graph.

For $J\subseteq V$, the subgraph induced by $J$ is
\begin{equation}
\gf[J]=(J,E[J]),
\qquad
E[J]=\{\{u,v\}\in E\mid u,v\in J\}.
\end{equation}
The neighborhood of $v\in V$ is
$N(v)=\{u\in V\mid \{u,v\}\in E\}$.

A subset of $V$ is \emph{independent} if no two of its vertices are
adjacent, and it is a \emph{clique} if every two distinct vertices in it
are adjacent.
We denote the collection of independent sets by $\mathcal I(\gf)$,
the collection of those of cardinality $k$ by $\mathcal I_k(\gf)$,
and the maximum cardinality of an independent set by $\alpha(\gf)$.

\subsection{Indifference graphs and their orderings}\label{ss:ig}

Let $v_1<\cdots<v_n$ be a total ordering of the vertices of $\gf$, and set
$N_i^-=N(v_i)\cap\{v_1,\ldots,v_{i-1}\}$ and
$N_i^+=N(v_i)\cap\{v_{i+1},\ldots,v_n\}$.
The ordering is called an \emph{indifference ordering} if
$N_i^-$ and $N_i^+$ are cliques for every $i$.
A graph admitting an indifference ordering is called an
\emph{indifference graph}.

\begin{figure}[htbp]
\centering
\begin{minipage}{.43\linewidth}
\centering
\begin{tikzpicture}[v/.style={circle,draw,inner sep=3pt}]
\node[v] (a) at (0,0) {$v_1$};
\node[v] (b) at (1.4,1) {$v_2$};
\node[v] (c) at (2.8,0) {$v_3$};
\node[v] (d) at (4.2,1) {$v_4$};
\foreach \i/\j in {a/b,a/c,b/c,b/d,c/d}{\draw (\i)--(\j);}
\end{tikzpicture}
\end{minipage}\hfill
\begin{minipage}{.53\linewidth}
\centering
\begin{tabular}{c|cc}
$i$ & $N_i^-$ & $N_i^+$\\\hline
$1$ & $\varnothing$ & $\{v_2,v_3\}$\\
$2$ & $\{v_1\}$ & $\{v_3,v_4\}$\\
$3$ & $\{v_1,v_2\}$ & $\{v_4\}$\\
$4$ & $\{v_2,v_3\}$ & $\varnothing$
\end{tabular}
\end{minipage}
\caption{An indifference ordering $v_1<v_2<v_3<v_4$.
Each listed neighbor set is a clique.}
\label{fig:ind}
\end{figure}
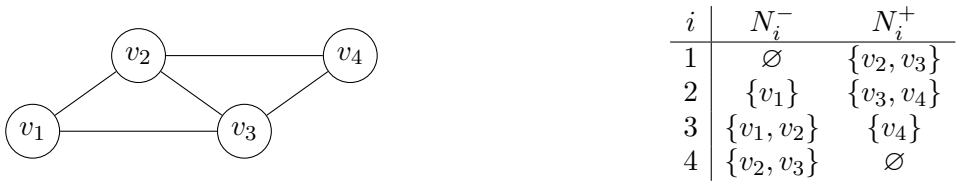

For example, the graph in Figure~\ref{fig:ind} has the
indifference ordering $v_1<v_2<v_3<v_4$.

The following characterization relates the local clique conditions
to the umbrella property, which underlies the interval representation
and path-power embedding below. For the ordering characterization of
indifference graphs, see also~\cite{LO93}.

\begin{prop}[Umbrella characterization]\label{pr:ub}
Let $v_1<\cdots<v_n$ be an ordering of the vertices of a connected
graph $\gf$.
Then the following conditions are equivalent:
\begin{enumerate}
\item
The ordering is an indifference ordering, namely, $N_i^-$ and
$N_i^+$ are cliques for every $i$.
\item
The ordering satisfies the umbrella property
\begin{equation}\label{up}
i<j<k,\qquad \{v_i,v_k\}\in E
\quad\Longrightarrow\quad
\{v_i,v_j\},\{v_j,v_k\}\in E.
\end{equation}
\end{enumerate}
\end{prop}

\begin{proof}
For a connected graph with at least two vertices, this equivalence is
stated in \cite[Remark~2.6]{CR12}.
The one-vertex case is immediate.
\end{proof}

\begin{remark}
For a disconnected graph, Proposition~\ref{pr:ub} may be applied
to each connected component separately.
Concatenating the resulting orderings gives an umbrella ordering of
the whole graph.
Thus, connected or not, a graph is an indifference graph if and only if
it admits an umbrella ordering.
\end{remark}

Once an indifference ordering is fixed, we orient every edge from its
earlier endpoint to its later endpoint and refer to the resulting
directed graph as an \emph{oriented indifference graph}.

\subsection{Path powers}\label{ss:pp}

Let $P_M$ be the graph on the ordered vertex set $\{1,\ldots,M\}$
whose edges join consecutive vertices.
For $p\ge1$, the $p$th path power $P_M^p$ is the graph on the same
vertex set in which two distinct vertices are adjacent if and only if
their distance in $P_M$ is at most $p$.
Equivalently,
\begin{equation}\label{pp}
\{\ell,m\}\in E(P_M^p)
\quad\Longleftrightarrow\quad
1\le |m-\ell|\le p.
\end{equation}
In particular, $P_M^1=P_M$.

We use the natural ordering and orient each edge from the smaller index
to the larger one.
Then
$N^-_\ell=\{\max(1,\ell-p),\ldots,\ell-1\}$ and
$N^+_\ell=\{\ell+1,\ldots,\min(M,\ell+p)\}$,
both of which are cliques.
Thus $P_M^p$ is an indifference graph.

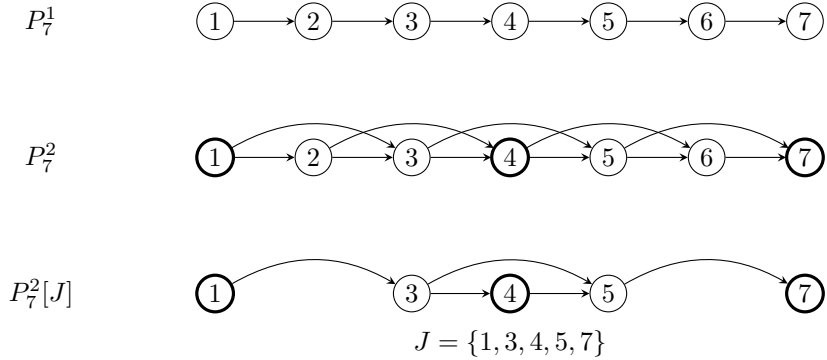
\begin{figure}[H]
\centering
\begin{tikzpicture}[
  v/.style={circle,draw,fill=white,inner sep=2pt},
  every node/.style={font=\small},
  >=stealth
]

\node at (-1.0,0) {$P_7^1$};
\foreach \i in {1,...,7}{
  \node[v] (a\i) at (1.3*\i,0) {$\i$};
}
\foreach \i in {1,...,6}{
  \pgfmathtruncatemacro{\j}{\i+1}
  \draw[->] (a\i) -- (a\j);
}

\node at (-1.0,-1.8) {$P_7^2$};
\foreach \i in {1,...,7}{
  \node[v] (b\i) at (1.3*\i,-1.8) {$\i$};
}
\foreach \i in {1,...,6}{
  \pgfmathtruncatemacro{\j}{\i+1}
  \draw[->] (b\i) -- (b\j);
}
\foreach \i in {1,...,5}{
  \pgfmathtruncatemacro{\j}{\i+2}
  \draw[->] (b\i) to[bend left=30] (b\j);
}
\foreach \i in {1,4,7}{
  \node[v,very thick] at (1.3*\i,-1.8) {$\i$};
}

\node at (-1.0,-3.6) {$P_7^2[J]$};

\node[v,very thick] (c1) at (1.3*1,-3.6) {$1$};
\node[v]            (c3) at (1.3*3,-3.6) {$3$};
\node[v,very thick] (c4) at (1.3*4,-3.6) {$4$};
\node[v]            (c5) at (1.3*5,-3.6) {$5$};
\node[v,very thick] (c7) at (1.3*7,-3.6) {$7$};

\draw[->] (c1) to[bend left=30] (c3);
\draw[->] (c3) -- (c4);
\draw[->] (c3) to[bend left=30] (c5);
\draw[->] (c4) -- (c5);
\draw[->] (c5) to[bend left=30] (c7);

\node[below] at (5.2,-3.95) {$J=\{1,3,4,5,7\}$};

\end{tikzpicture}
\caption{
The path powers $P_7^1$ and $P_7^2$, together with the induced
subgraph $P_7^2[J]$ for $J=\{1,3,4,5,7\}$, with edges oriented from
smaller to larger vertices. The vertices $\{1,4,7\}$ form an
independent set in both $P_7^2$ and $P_7^2[J]$.
All three graphs are indifference graphs.}
\label{f:pp}
\end{figure}

For later use, we write
\begin{equation}\label{prec}
i\prec j
\quad\Longleftrightarrow\quad
i+p<j.
\end{equation}
Thus two vertices $i<j$ of $P_M^p$ are nonadjacent precisely when
$i\prec j$.
In particular,
$I=\{i_1<\cdots<i_k\}$ belongs to $\mathcal I(P_M^p)$ if and only if
$i_1\prec\cdots\prec i_k$.
More generally, for $J\subseteq\{1,\ldots,M\}$,
$I=\{i_1<\cdots<i_k\}$ belongs to $\mathcal I(P_M^p[J])$ precisely
when $i_r\in J$ for all $r$ and $i_1\prec\cdots\prec i_k$.
See Figure~\ref{f:pp} for examples.

\subsection{Embedding into a path power}\label{ss:epp}

We first recall the realization of finite indifference graphs
as induced subgraphs of powers of paths in the form needed below.
An interval representation assigns a closed interval to each vertex,
with adjacency equivalent to intersection. It is called \emph{proper}
if no interval properly contains another, and \emph{unit} if all
intervals have the same positive length, normalized here to $1$.

The following proposition is essentially a consequence of known results,
but we record the order-preserving form needed for the Q6V construction.

\begin{prop}\label{prp}
Let $\gf$ be a nonempty finite indifference graph with an indifference
ordering $v_1<\cdots<v_n$ in which every connected component forms a
consecutive block.
Then there exist a positive integer $p$ and integers
$1=m_1<\cdots<m_n$ such that
\begin{equation}\label{emb}
\{v_i,v_j\}\in E(\gf)
\quad\Longleftrightarrow\quad
1\le m_j-m_i\le p
\qquad (1\le i<j\le n).
\end{equation}
Consequently, for every $M\ge m_n$, the map $v_i\mapsto m_i$ is an
order-preserving graph isomorphism from $\gf$ onto the induced subgraph
$P_M^p[\{m_1,\ldots,m_n\}]$ of the path power $P_M^p$.
\end{prop}

\begin{proof}
A proper interval representation respecting the given order within
each connected component exists
\cite[Theorem~2.3 and Remark~2.6]{CR12}.
Such a representation can be converted to a unit interval representation
while preserving the endpoint order \cite{BW99}.
Such a unit interval representation yields an order-preserving
induced embedding into a suitable path power
\cite[proof of Theorem~9(ii), pp.~626--627]{LRSS11}.
Thus we obtain $v_i\mapsto m_i$ into some $P_M^p$.
After translating the indices, we may take $m_1=1$,
which gives \eqref{emb}.
\end{proof}

\begin{remark}
For a fixed integer embedding $m_1<\cdots<m_n$, the value of $p$
need not be unique.
If both edges and nonedges are present, every integer $p$ satisfying
\begin{equation}
\max_{\substack{\{v_i,v_j\}\in E(\gf)\\ i<j}}(m_j-m_i)
\le p
<
\min_{\substack{\{v_i,v_j\}\notin E(\gf)\\ i<j}}(m_j-m_i)
\end{equation}
gives the same induced graph on $\{m_1,\ldots,m_n\}$.
If one of the two sets is empty, the corresponding one-sided condition
is understood.
The value $p=D$ arising from the explicit construction in
Appendix~\ref{app:embed} is simply one convenient admissible choice.
No minimality or uniqueness of $p$ will be needed below.
\end{remark}

A related description in terms of the non-homogeneous interaction
ranges considered in~\cite{A25} is given in Appendix~\ref{app:nhr}.

\subsection{Example: a triangle with a pendant vertex}\label{ss:exp}

Let $\gf$ have vertices $v_1<v_2<v_3<v_4$ and edge set
$E(\gf)=\{\{v_1,v_2\},\{v_1,v_3\},\{v_2,v_3\},\{v_3,v_4\}\}$.
Thus vertices $v_1,v_2,v_3$ form a triangle, and vertex $v_4$ is adjacent
only to vertex $v_3$.
This is an indifference ordering: the only neighbor sets with more
than one vertex are $N_1^+=\{v_2,v_3\}$ and $N_3^-=\{v_1,v_2\}$, both of
which are cliques.

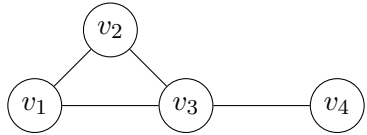
\begin{figure}[htbp]
\centering
\begin{tikzpicture}[v/.style={circle,draw,inner sep=3pt}]
\node[v] (v1) at (0,0) {$v_1$};
\node[v] (v2) at (1,1) {$v_2$};
\node[v] (v3) at (2,0) {$v_3$};
\node[v] (v4) at (4,0) {$v_4$};
\foreach \i/\j in {1/2,1/3,2/3,3/4}
  {\draw (v\i)--(v\j);}
\end{tikzpicture}
\caption{A triangle with a pendant vertex, in the ordering
$v_1<v_2<v_3<v_4$.}
\label{fex1}
\end{figure}

This graph is not itself a power of a path on four vertices.
It is not $P_4$, since it contains a triangle.
In $P_4^2$ there are two vertices adjacent to all three other vertices,
whereas $\gf$ has only one.
Finally, the entire vertex set of $P_4^p$ is a clique for $p\ge3$,
whereas that of $\gf$ is not.

For this ordering, the indices $\ell_i$ defined by \eqref{eli} are
$(\ell_1,\ell_2,\ell_3,\ell_4)=(1,1,1,3)$.
Applying the recursive construction \eqref{app:xrec} in
Appendix~\ref{app:embed} then gives
\begin{equation}
(x_1,x_2,x_3,x_4)=(0,\tfrac12,\tfrac34,\tfrac{13}{8}).
\end{equation}
The first three coordinates lie within distance $1$ of one another,
while
$x_4-x_3=7/8<1$, $x_4-x_2=9/8>1$, and
$x_4-x_1=13/8>1$.
Hence the intervals $[x_i,x_i+1]$ represent precisely the edges of
$\gf$.

\begin{figure}[htbp]
\centering
\begin{tikzpicture}[every node/.style={font=\small}]
\foreach \y/\x/\n in
  {0/0/v_1,-.5/1/v_2,-1/1.5/v_3,-1.5/3.25/v_4}{
  \draw[|-|,thick] (\x,\y)--(\x+2,\y);
  \node[left] at (\x,\y) {$\n$};
}
\end{tikzpicture}
\caption{A unit interval representation with left endpoints
$(0,\frac12,\frac34,\frac{13}{8})$.}
\label{fex2}
\end{figure}
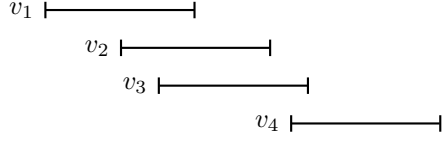

Clearing the denominators with $D=p=8$ gives
$(m_1,m_2,m_3,m_4)=(1,5,7,14)$.
Consequently,
\begin{equation}
\gf\simeq P_{14}^{8}[\{1,5,7,14\}].
\end{equation}
Indeed, the differences corresponding to the four edges are
$4,6,2,7$, all at most $8$, whereas those corresponding to the
two nonedges are $13$ and $9$, both greater than $8$.

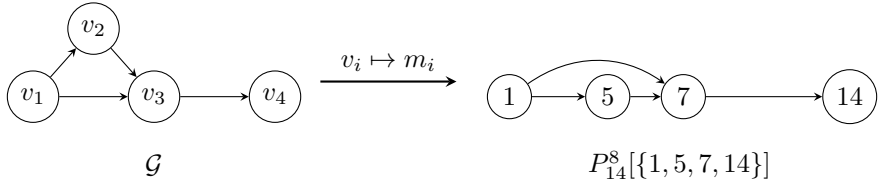
\begin{figure}[htbp]
\centering
\begin{tikzpicture}[>=stealth,
                    v/.style={circle,draw,inner sep=3pt},
                    every node/.style={font=\small}]
\node[v] (a1) at (0,0) {$v_1$};
\node[v] (a2) at (.8,.9) {$v_2$};
\node[v] (a3) at (1.6,0) {$v_3$};
\node[v] (a4) at (3.2,0) {$v_4$};
\foreach \i/\j in {1/2,1/3,2/3,3/4}
  {\draw[->] (a\i)--(a\j);}
\node at (1.6,-.9) {$\gf$};

\draw[->,thick] (3.8,.2)--(5.6,.2)
  node[midway,above] {$v_i\mapsto m_i$};

\begin{scope}[xshift=6.3cm]
  \foreach \x/\i in {0/1,1.3/5,2.3/7,4.5/14}
    {\node[v] (b\i) at (\x,0) {$\i$};}
  \foreach \i/\j in {1/5,5/7,7/14}
    {\draw[->] (b\i)--(b\j);}
  \draw[->] (b1) to[bend left=35] (b7);
  \node at (2.25,-.9) {$P_{14}^{8}[\{1,5,7,14\}]$};
\end{scope}
\end{tikzpicture}
\caption{The induced embedding of the oriented graph $\gf$
into $P_{14}^{8}$, with every edge directed from the smaller
vertex to the larger one.
Only the relevant vertices and the edges among them are shown.}
\label{fex3}
\end{figure}

For later reference, the independent sets of this graph are
\begin{equation}
\mathcal I(\gf)
=
\bigl\{
\varnothing,
\{v_1\},\{v_2\},\{v_3\},\{v_4\},
\{v_1,v_4\},\{v_2,v_4\}
\bigr\}.
\end{equation}

\subsection{Independent-set operators and their properties}
\label{ss:quo}

We recall the operator identities associated with an oriented
indifference graph; see \cite{MEWC25}.
Let $\gf=(V,E)$ be such a graph, with each edge directed from the
earlier vertex to the later one in an indifference ordering.
Fix an integer $N\ge2$ and a primitive $N$th root of unity $\omega$.

Associate an operator $h_i$ with each vertex $i\in V$ and assume
\begin{equation}\label{halg}
h_i h_j=
\begin{cases}
\omega h_j h_i, & i\longrightarrow j,\\
h_j h_i, & \{i,j\}\notin E,
\end{cases}
\qquad
h_i^N=b_i^N\mathrm{Id},
\end{equation}
where $b_i$ is a scalar and the first case refers to a directed edge
of $\gf$.
We refer to the oriented graph specifying these exchange relations as
the \emph{frustration graph}.
Its vertices label interaction terms rather than physical spin sites.
The scalar $b_i$ may be zero.

For $I\in\mathcal I(\gf)$, put
$h_I=\prod_{i\in I}h_i$ and $b_I=\prod_{i\in I}b_i$, with
$h_{\varnothing}=\mathrm{Id}$ and $b_{\varnothing}=1$.
The factors in $h_I$ commute, so their order is immaterial.

Define the independent-set operators
\begin{equation}\label{tran}
\mathcal T_{\gf}(y)
=
\sum_{I\in\mathcal I(\gf)}y^{|I|}h_I
=
\sum_{k=0}^{\alpha(\gf)}y^k\mathcal Q_k,
\qquad
\mathcal Q_k=
\sum_{I\in\mathcal I_k(\gf)}h_I.
\end{equation}
In particular, $\mathcal Q_0=\mathrm{Id}$ and
$\mathcal Q_1=\sum_{i\in V}h_i$ is referred to as the Hamiltonian.
We set 
\begin{equation}\label{zgn}
\mathcal Z_{\gf,N}(y)=
\sum_{I\in\mathcal I(\gf)}y^{|I|}b_I^N.
\end{equation}

\begin{thm}[\cite{MEWC25}]\label{th:g}
Under the assumptions above,
\begin{align}
[\mathcal T_{\gf}(y),\mathcal T_{\gf}(y')]
&=0,
\label{tcom}
\\
\mathcal T_{\gf}(y)\mathcal T_{\gf}(\omega y)
\cdots
\mathcal T_{\gf}(\omega^{N-1}y)
&=
\mathcal Z_{\gf,N}((-1)^{N+1}y^N)\,\mathrm{Id}.
\label{frob}
\end{align}
In particular, the operators $\mathcal Q_k$ commute with one another.
\end{thm}

For odd $N$, Theorem~\ref{th:main} below realizes the system
\eqref{halg} by the root-of-unity Q6V model and identifies
$\mathcal T_{\gf}(y)$ with a specialized Q6V transfer matrix.
Consequently, Theorem~\ref{th:g} can also be recovered from the
corresponding Q6V results of \cite{IKTY25,IKTY26}.

We note that the assertions in Theorem~\ref{th:g} were shown by direct
manipulations for concrete models corresponding to path-power
frustration graphs $\gf=P_M^p$ for some small $p$ in
\cite{AP2,F14,F19}.

\section{Quantized six-vertex model}\label{s:Q6V}

In this section, we formulate the Q6V model on the wiring diagram
$G_{p,L}$ and describe its states and weights.
We then introduce the elementary paths and show that their compatibility
is governed by the path power $P_M^p$.
After specializing the parameters, the Q6V states of nonzero weight are
put in one-to-one correspondence with independent sets of a suitable
induced subgraph of $P_M^p$.
As a consequence, the Q6V transfer matrix is expressed as an
independent-set generating function.

\subsection{Local weights}\label{ss:loc}

\begin{figure}[H]
\centering
{\unitlength 0.011in
\begin{picture}(525,75)(-15,30)
\put(6,80){
  \put(-11,0){\vector(1,0){23}}
  \put(0,-10){\vector(0,1){22}}
}
\multiput(81,80.5)(75,0){6}{
  \put(-11,0){\vector(1,0){23}}
  \put(0,-10){\vector(0,1){22}}
}
\put(-74,0){
  \put(60.5,77){$i$}
  \put(77.5,60){$j$}
  \put(96,77){$a$}
  \put(77.5,96.5){$b$}
}
\put(61,77){0}
\put(78,60){0}
\put(96,77){0}
\put(78,96.5){0}
\put(75,0){
  \put(61,77){1}
  \put(78,60){1}
  \put(96,77){1}
  \put(78,96.5){1}
}
\put(150,0){
  \put(61,77){1}
  \put(78,60){0}
  \put(96,77){1}
  \put(78,96.5){0}
}
\put(225,0){
  \put(61,77){0}
  \put(78,60){1}
  \put(96,77){0}
  \put(78,96.5){1}
}
\put(300,0){
  \put(61,77){0}
  \put(78,60){1}
  \put(96,77){1}
  \put(78,96.5){0}
}
\put(375,0){
  \put(61,77){1}
  \put(78,60){0}
  \put(96,77){0}
  \put(78,96.5){1}
}

\put(78,43){
  \put(-77,0){$\mathcal L^{ab}_{ij}$}
  \put(0,0){$r$}
  \put(75,0){$s$}
  \put(144,0){$fU$}
  \put(222,0){$gU$}
  \put(291,0){$W^{-1}$}
  \put(340,0){$rsW+ qfg WU^2$}
}
\end{picture}
}
\caption{The index convention (left) and the six nonzero
$\mathcal L^{ab}_{ij}$.}
\label{f:loc}
\end{figure}
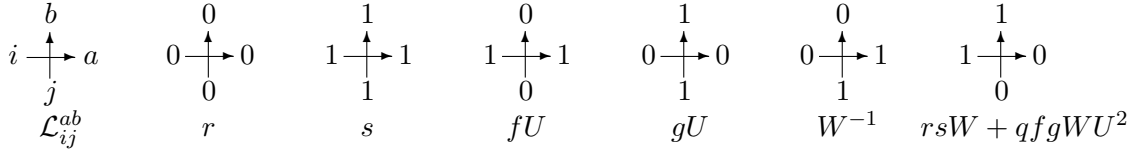

Let $\mathcal W(q)$ be the $q$-Weyl algebra generated by
$U^{\pm1},W^{\pm1}$ subject to
\begin{equation}\label{uw}
UW=qWU.
\end{equation}
We assume that $q$ is generic in this section.
For $a,b,i,j\in\{0,1\}$, define the element 
$\LL^{ab}_{ij}=\LL^{ab}_{ij}(r,s,f,g;U,W)\in\mathcal W(q)$
by the conservation law
\begin{equation}\label{wp}
\LL^{ab}_{ij}=0\qquad\text{unless}\qquad a+b=i+j,
\end{equation}
and
\begin{equation}\label{les}
\LL^{00}_{00}=r,\quad
\LL^{11}_{11}=s,\quad
\LL^{10}_{10}=fU,\quad
\LL^{01}_{01}=gU,\quad
\LL^{10}_{01}=W^{-1},\quad
\LL^{01}_{10}=rsW+qfgWU^2.
\end{equation}
We depict them as in Figure~\ref{f:loc}.

The definitions \eqref{les} and Figure~\ref{f:loc} are taken from
\cite[(2.3c)]{IKTY26} and \cite[Fig.~1]{IKTY26} by setting
$U=e^\uu$, $W=e^\ww$ and applying the Baker--Campbell--Hausdorff formula.

\subsection{Q6V model on $G_{p,L}$}\label{ss:vG}

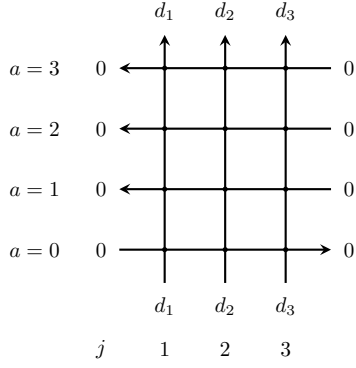
\begin{figure}[H]
\centering
\begin{tikzpicture}[>=stealth,scale=.8,transform shape,
  every node/.style={font=\small}]

  \foreach \x/\j in {1/1,2/2,3/3}{
    \draw[->,thick] (\x,-.55)--(\x,3.55);
    \node[above=3pt] at (\x,3.55) {$d_{\j}$};
    \node[below=3pt] at (\x,-.55) {$d_{\j}$};
    \node at (\x,-1.65) {$\j$};
  }

  \foreach \a in {1,2,3}{
    \draw[<-,thick] (.25,\a)--(3.75,\a);
    \node at (-1.15,\a) {$a=\a$};
    \node at (-.05,\a) {$0$};
    \node at (4.05,\a) {$0$};
  }

  \draw[->,thick] (.25,0)--(3.75,0);
  \node at (-1.15,0) {$a=0$};
  \node at (-.05,0) {$0$};
  \node at (4.05,0) {$0$};

  \foreach \x in {1,2,3}
    \foreach \a in {0,1,2,3}
      \fill (\x,\a) circle (1.2pt);

  \node at (-.05,-1.65) {$j$};
\end{tikzpicture}
\caption{An example of the admissible wiring diagram $G_{p,L}$ consisting of 
the rows $a=0,1,\ldots, p$ and the columns $j=1,\ldots, L$ for $p=3$ and $L=3$.
Horizontal boundary states are fixed to $0$ and
the top and bottom boundary states are required to coincide in each
column $j$, and we denote their common value by $d_j\in\{0,1\}$.}
\label{f:grid}
\end{figure}

Let $p,L\ge1$ be integers.
We formulate the Q6V model on the wiring diagram $G=G_{p,L}$
consisting of $p+1$ horizontal and $L$ vertical directed lines
arranged in a rectangular grid.
The horizontal lines are labeled $a=0,\ldots,p$ from bottom to top,
and the vertical lines are labeled $j=1,\ldots,L$ from left to right.
The bottom horizontal line points to the right, the remaining
horizontal lines point to the left, and all vertical lines point upward.
The Q6V model is defined on admissible wiring diagrams in the sense of
\cite[Def.~3.3]{IKTY25} and \cite[Def.~2.3]{IKTY26},
and $G_{p,L}$ satisfies this admissibility condition.
It has $K=(p+1)L$ vertices.
The vertex at the intersection of column $j$ and row $a$ is denoted by $(j,a)$.
An example is shown in Figure~\ref{f:grid}.

Each edge of $G$ carries a local state $0$ or $1$.
We call an edge with state $1$ \emph{occupied} and one with state $0$
\emph{unoccupied}.
We impose fixed boundary conditions horizontally and periodic boundary
conditions vertically.
All $2(p+1)$ horizontal boundary states at the left and right ends are
fixed to $0$.
In each column $j$, the top and bottom boundary states are required to
coincide; their common value is denoted by $d_j\in\{0,1\}$.
These boundary conditions are indicated in Figure~\ref{f:grid}.

A state $\st$ of the Q6V model is an assignment of local states $0$ or $1$
to all edges of $G$ satisfying the boundary conditions above and the
conservation law \eqref{wp} at every vertex.
We denote by $\cc=\cc(G)$ the set of all such states.
For $\st\in\cc$, define
\begin{subequations}\label{ws}
\begin{align}
\wt(\st)
&=\bigotimes_{v=1}^K
\LL^{a_v,b_v}_{i_v,j_v}(r_v,s_v,f_v,g_v;U_v,W_v)
\in\mathcal W(q)^{\otimes K},
\label{wsw}\\
\dd(\st)&=d_1+\cdots+d_L.
\label{wsd}
\end{align}
\end{subequations}
Thus $\dd(\st)$ is the number of occupied vertical boundary edges,
which we call the \emph{degree} of the state $\st$.
The product in \eqref{wsw} extends over all vertices of $G$, with each
factor determined from $\st$ according to Figure~\ref{f:loc}.
The operators $U_v,W_v$ are independent copies of the $q$-Weyl algebra
generators and satisfy
\begin{equation}\label{uwv}
U_vW_{v'}=q^{\delta_{v,v'}}W_{v'}U_v.
\end{equation}
Hence the ordering in \eqref{wsw} is immaterial.

The layer transfer matrix is defined by
\begin{equation}\label{TG}
T_G(y)=\sum_{\st\in\cc}y^{\dd(\st)}\wt(\st).
\end{equation}
For generic parameters, it is a polynomial in $y$ of degree $L$
with coefficients in $\mathcal W(q)^{\otimes K}$.
The factor $y^{\dd(\st)}$ may be regarded as a boundary fugacity.
The dependence of $T_G(y)$ on the local parameters is suppressed
from the notation.
The tetrahedron equations and inversion relations for the operators
$\LL^{ab}_{ij}$ \cite[Prop.~2.1]{IKTY26} imply
$[T_G(y),T_G(z)]=0$; see \cite[Prop.~2.8]{IKTY26}.

\subsection{Five-vertex specialization}\label{ss:spe}

We now specialize the parameters other than $q$.
At the vertex $(j,a)$, write the parameters and $q$-Weyl generators in
\eqref{wsw} as
$(r_{j,a},s_{j,a},f_{j,a},g_{j,a})$ and $U_{j,a},W_{j,a}$.
For $1\le j\le L$, we take
\begin{equation}\label{spep}
\begin{array}{c|cccc}
$a$ & r_{j,a} & s_{j,a} & f_{j,a} & g_{j,a} \\ \hline
1,\ldots,p & 1 & s_{j,a} & 1 & 0 \\
0           & 1 & 1       & 0 & g_{j,0}.
\end{array}
\end{equation}

\begin{figure}[H]
\centering
\begin{tikzpicture}[
  thinwire/.style={line width=.34pt},
  pathwire/.style={
    line width=1.65pt,
    line cap=round,
    line join=round
  }
]

\newcommand{\localN}[8]{%
\begin{scope}[shift={(#1,#2)}]

  \begin{scope}[shift={(.10,0)},x=.8cm,y=.8cm]

    \node at (-.70,0) {$#4$};
    \node at (0,-.74) {$#5$};
    \node at (.70,0) {$#6$};
    \node at (0,.74) {$#7$};

    \ifnum#6=1
      \draw[thinwire] (-.46,0)--(.46,0);
    \else
      \draw[
        thinwire,
        -{Latex[length=1.7mm,width=1.25mm]}
      ] (-.46,0)--(.46,0);
    \fi

    \ifnum#7=1
      \draw[thinwire] (0,-.46)--(0,.46);
    \else
      \draw[
        thinwire,
        -{Latex[length=1.7mm,width=1.25mm]}
      ] (0,-.46)--(0,.46);
    \fi

    \def\typ{#3}

    \def\tmp{X}
    \ifx\typ\tmp
      \draw[
        pathwire,
        -{Latex[length=2.0mm,width=1.45mm]}
      ] (-.46,0)--(.46,0);
      \draw[
        pathwire,
        -{Latex[length=2.0mm,width=1.45mm]}
      ] (0,-.46)--(0,.46);
    \fi

    \def\tmp{H}
    \ifx\typ\tmp
      \draw[
        pathwire,
        -{Latex[length=2.0mm,width=1.45mm]}
      ] (-.46,0)--(.46,0);
    \fi

    \def\tmp{V}
    \ifx\typ\tmp
      \draw[
        pathwire,
        -{Latex[length=2.0mm,width=1.45mm]}
      ] (0,-.46)--(0,.46);
    \fi

    \def\tmp{BR}
    \ifx\typ\tmp
      \draw[
        pathwire,
        -{Latex[length=2.0mm,width=1.45mm]}
      ] (0,-.46)--(0,0)--(.46,0);
    \fi

    \def\tmp{LT}
    \ifx\typ\tmp
      \draw[
        pathwire,
        -{Latex[length=2.0mm,width=1.45mm]}
      ] (-.46,0)--(0,0)--(0,.46);
    \fi

  \end{scope}

  \node at (.10,-1.15) {$#8$};

\end{scope}
}

\newcommand{\localR}[8]{%
\begin{scope}[shift={(#1,#2)}]

  \begin{scope}[shift={(.10,0)},x=.8cm,y=.8cm]

    \node at (-.70,0) {$#4$};
    \node at (0,-.74) {$#5$};
    \node at (.70,0) {$#6$};
    \node at (0,.74) {$#7$};

    \ifnum#4=1
      \draw[thinwire] (.46,0)--(-.46,0);
    \else
      \draw[
        thinwire,
        -{Latex[length=1.7mm,width=1.25mm]}
      ] (.46,0)--(-.46,0);
    \fi

    \ifnum#7=1
      \draw[thinwire] (0,-.46)--(0,.46);
    \else
      \draw[
        thinwire,
        -{Latex[length=1.7mm,width=1.25mm]}
      ] (0,-.46)--(0,.46);
    \fi

    \def\typ{#3}

    \def\tmp{X}
    \ifx\typ\tmp
      \draw[
        pathwire,
        -{Latex[length=2.0mm,width=1.45mm]}
      ] (.46,0)--(-.46,0);
      \draw[
        pathwire,
        -{Latex[length=2.0mm,width=1.45mm]}
      ] (0,-.46)--(0,.46);
    \fi

    \def\tmp{H}
    \ifx\typ\tmp
      \draw[
        pathwire,
        -{Latex[length=2.0mm,width=1.45mm]}
      ] (.46,0)--(-.46,0);
    \fi

    \def\tmp{V}
    \ifx\typ\tmp
      \draw[
        pathwire,
        -{Latex[length=2.0mm,width=1.45mm]}
      ] (0,-.46)--(0,.46);
    \fi

    \def\tmp{RT}
    \ifx\typ\tmp
      \draw[
        pathwire,
        -{Latex[length=2.0mm,width=1.45mm]}
      ] (.46,0)--(0,0)--(0,.46);
    \fi

    \def\tmp{LB}
    \ifx\typ\tmp
      \draw[
        pathwire,
        -{Latex[length=2.0mm,width=1.45mm]}
      ] (0,-.46)--(0,0)--(-.46,0);
    \fi

  \end{scope}

  \node at (.10,-1.15) {$#8$};

\end{scope}
}


\localR{0.00}{1.45}{Z}  {0}{0}{0}{0}{1}
\localR{2.35}{1.45}{X}  {1}{1}{1}{1}{s_{j,a}}
\localR{4.70}{1.45}{V}  {0}{1}{0}{1}{U_{j,a}}
\localR{7.05}{1.45}{RT} {0}{0}{1}{1}{W^{-1}_{j,a}}
\localR{9.40}{1.45}{LB} {1}{1}{0}{0}{s_{j,a}W_{j,a}}


\localN{0.00}{-1.30}{Z}  {0}{0}{0}{0}{1}
\localN{2.35}{-1.30}{X}  {1}{1}{1}{1}{1}
\localN{4.70}{-1.30}{V}  {0}{1}{0}{1}{g_{j,0}U_{j,0}}
\localN{7.05}{-1.30}{BR} {0}{1}{1}{0}{W^{-1}_{j,0}}
\localN{9.40}{-1.30}{LT} {1}{0}{0}{1}{W_{j,0}}

\end{tikzpicture}

\caption{The nonzero local weights under the five-vertex specialization \eqref{spep}.
Upper and lower rows apply to the rows $a=1,\ldots,p$ and $a=0$
of $G_{p,L}$ in Figure~\ref{f:grid},
respectively. In the former, the vertices are rotated 90 degrees
anticlockwise, reflecting the orientation in Figure~\ref{f:grid}.
In either row, the nonzero local configurations thus form a quantized
\emph{five}-vertex model.}
\label{f:los}
\end{figure}
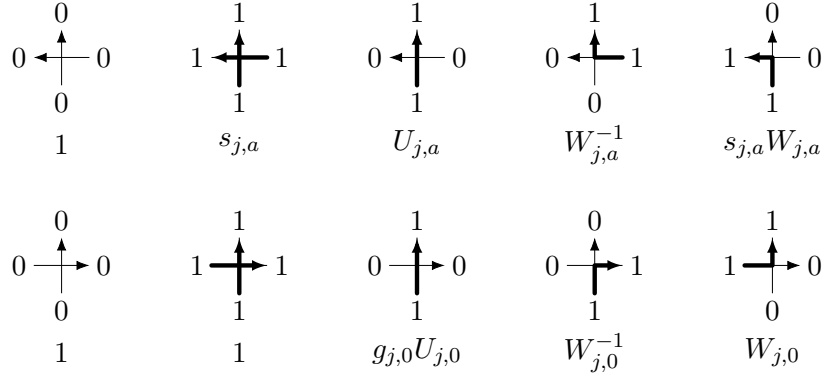

We represent a state graphically by drawing each occupied edge as a
solid line. Under \eqref{spep}, the nonzero local configurations are
those in Figure~\ref{f:los}.

\begin{figure}[H]
\centering
\begin{tikzpicture}[>={Latex[length=1.7mm,width=1.25mm]}]
\newcommand{\pathpanel}[5]{%
  \begin{scope}[shift={(#1,#2)},x=.42cm,y=.42cm,
    thinwire/.style={line width=.34pt},
    pathwire/.style={
      line width=1.65pt,
      line cap=round,
      line join=round
    }]
    \foreach \x in {1,2,3}
      \draw[thinwire,->] (\x,-.60)--(\x,3.90);

    \draw[thinwire,->] (.10,0)--(3.75,0);

    \foreach \y in {1,2,3}
      \draw[thinwire,<-] (.10,\y)--(3.75,\y);

    \foreach \x in {1,2,3}
      \foreach \y in {0,1,2,3}
        \fill (\x,\y) circle (1.05pt);

    \ifx#3A
      \draw[
        pathwire,
        -{Latex[length=2.0mm,width=1.45mm]}
      ]
        (#4,-.60)--(#4,3.90);
    \else
      \draw[pathwire]
        (#4,-.60)
        --(#4,0)
        --({#4+1},0)
        --({#4+1},#5)
        --(#4,#5);

      \draw[
        pathwire,
        -{Latex[length=2.0mm,width=1.45mm]}
      ]
        (#4,#5)--(#4,3.90);
    \fi
  \end{scope}%
}

\pathpanel{0}{3.00}{A}{1}{0}
\pathpanel{2.25}{3.00}{B}{1}{1}
\pathpanel{4.50}{3.00}{B}{1}{2}
\pathpanel{6.75}{3.00}{B}{1}{3}

\node at (.84,2.35) {$\hf_1=A_1$};
\node at (3.09,2.35) {$\hf_2=B_{1,1}$};
\node at (5.34,2.35) {$\hf_3=B_{1,2}$};
\node at (7.59,2.35) {$\hf_4=B_{1,3}$};

\pathpanel{0}{0}{A}{2}{0}
\pathpanel{2.25}{0}{B}{2}{1}
\pathpanel{4.50}{0}{B}{2}{2}
\pathpanel{6.75}{0}{B}{2}{3}
\pathpanel{9.00}{0}{A}{3}{0}

\node at (.84,-.65) {$\hf_5=A_2$};
\node at (3.09,-.65) {$\hf_6=B_{2,1}$};
\node at (5.34,-.65) {$\hf_7=B_{2,2}$};
\node at (7.59,-.65) {$\hf_8=B_{2,3}$};
\node at (9.84,-.65) {$\hf_9=A_3$};
\end{tikzpicture}
\caption{The nine states on $G_{3,3}$ with $\dd(\st)=1$
whose weights remain nonzero under the five-vertex specialization \eqref{spep},
depicted by their occupied edges.}
\label{fig:p9}
\end{figure}
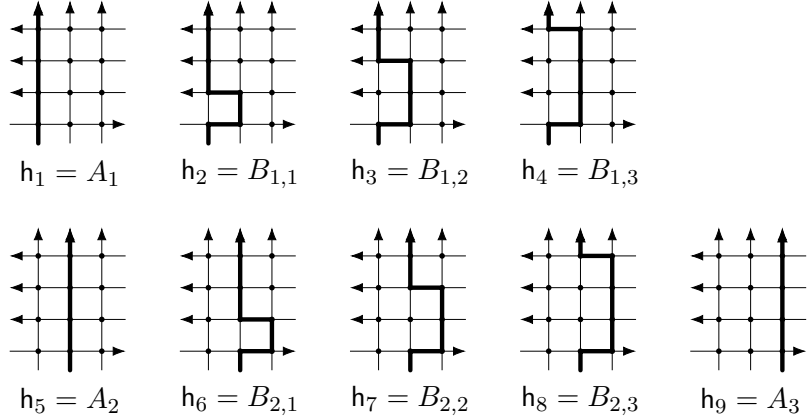

For generic nonzero values of $g_{j,0}$ and $s_{j,a}$, the number of
states of nonzero weight with $\dd(\st)=1$ is
\begin{equation}\label{M}
M=(p+1)(L-1)+1.
\end{equation}
Their occupied-edge configurations are shown in Figure~\ref{fig:p9}
for $p=L=3$.

In general, the elementary paths corresponding to states of nonzero
weight with $\dd(\st)=1$ and their weights
$\hf_1,\ldots,\hf_M$ are given by
\begin{subequations}\label{hs}
\begin{align}
\hf_{(p+1)(j-1)+1}
&=A_j
\;(1\le j\le L),
\quad
\hf_{(p+1)(j-1)+a+1}
=B_{j,a}
\; (1\le j<L,\ 1\le a\le p),
\\
A_j
&=
g_{j,0}U_{j,0}\cdots U_{j,p},
\\
B_{j,a}
&=
s_{j+1,a}
W^{-1}_{j,0}W_{j+1,0}
U_{j+1,1}\cdots U_{j+1,a-1}
W_{j+1,a}W^{-1}_{j,a}
U_{j,a+1}\cdots U_{j,p}.
\end{align}
\end{subequations}
We use the font $\hf_i$ to distinguish these elementary path weights
from the operators $h_i$ in \eqref{halg}.

\begin{lem}\label{le:halg}
The elementary path weights in \eqref{hs} satisfy
\begin{equation}\label{halg2}
\hf_i\hf_j=
\begin{cases}
q^{-2}\hf_j\hf_i,
&i<j\le i+p,
\\
\hf_j\hf_i,
&i+p<j.
\end{cases}
\end{equation}
\end{lem}

\begin{proof}
The relations can be verified directly case by case, as illustrated
by Figure~\ref{fig:p9}.
The relevant cases are
\begin{equation}
\begin{array}{c@{\qquad\qquad}c}
\text{\rm Representative case in Figure~\ref{fig:p9}}
&
\!\!\!\!\!\!\!\!\!\!\!\!\!\!\!\!\!\!\!\!\!\!\!\!\!\!\!\!
\!\!\!\!\!\!\!\!\!\!\!\!
\text{\rm General relation}
\\[1mm]
\begin{aligned}
\hf_1\hf_2 &= q^{-2}\hf_2\hf_1,\\
\hf_3\hf_5 &= q^{-2}\hf_5\hf_3,\\
\hf_2\hf_3 &= q^{-2}\hf_3\hf_2,\\
\hf_4\hf_6 &= q^{-2}\hf_6\hf_4,\\
\hf_3\hf_7 &= \hf_7\hf_3,
\end{aligned}
&
\begin{aligned}
A_jB_{j,a}
&=q^{-2}B_{j,a}A_j,\\
B_{j,a}A_{j+1}
&=q^{-2}A_{j+1}B_{j,a},\\
B_{j,a}B_{j,b}
&=q^{-2}B_{j,b}B_{j,a}
\qquad(a<b),\\
B_{j,a}B_{j+1,b}
&=q^{-2}B_{j+1,b}B_{j,a}
\qquad(a>b),\\
B_{j,a}B_{j+1,b}
&=B_{j+1,b}B_{j,a}
\qquad(a\le b).
\end{aligned}
\end{array}
\end{equation}
All other pairs commute.
For example, the relation
$\hf_2\hf_3=q^{-2}\hf_3\hf_2$
follows directly from
\begin{equation}
\begin{split}
\hf_2 &= B_{1,1} = 
s_{2,1}
W^{-1}_{1,0}W_{2,0}
W_{2,1}W^{-1}_{1,1}
U_{1,2}U_{1,3},
\\
\hf_3 &= B_{1,2} = 
s_{2,2}
W^{-1}_{1,0}W_{2,0}
U_{2,1}
W_{2,2}W^{-1}_{1,2}
U_{1,3}.
\end{split}
\end{equation}
The only noncommuting factors occur at the sites
$(2,1)$ and $(1,2)$:
\begin{equation}
W_{2,1}U_{2,1}=q^{-1}U_{2,1}W_{2,1},
\qquad
U_{1,2}W^{-1}_{1,2}=q^{-1}W^{-1}_{1,2}U_{1,2}.
\end{equation}
Geometrically, the noncommutativity arises when the straight segment
of one elementary path meets a turning segment of another at a common
vertex. Each such encounter contributes a factor of $q^{\pm1}$.
\end{proof}

\subsection{States and independent sets}\label{ss:si}

For each elementary path weight $\hf_i$, let $c_i$ denote its scalar
coefficient. Thus, by \eqref{hs}, we have
\begin{subequations}
\begin{align}
&\hf_i = 
c_i\times
\bigl(\text{a Laurent monomial in the $U_{j,a}$ and $W_{j,a}$}\bigr),
\label{hcu}
\\
&c_{(p+1)(j-1)+1}=g_{j,0}
\;\, (1\le j\le L),
\quad\;
c_{(p+1)(j-1)+a+1}=s_{j+1,a}
\;\,(1\le j<L,\ 1\le a\le p).
\label{cdef}
\end{align}
\end{subequations}
The parameters $c_1,\ldots,c_M$ are mutually independent, since the
parameters on the right-hand side of \eqref{cdef} are independent.
When it is useful to display this dependence explicitly, we write
$T_G(y;\boldsymbol c)$ for \eqref{TG} and
$\wt_{\boldsymbol c}(\st)$ for \eqref{wsw}, where
$\boldsymbol c=(c_1,\ldots,c_M)$.
Set
\begin{equation}\label{jc}
J(\boldsymbol c)
= \{i\in\{1,\ldots,M\}\mid c_i\ne0\}.
\end{equation}

\begin{prop}\label{pr:Tg}
For the specialized Q6V model on $G=G_{p,L}$ and a fixed
$\boldsymbol c$, there is a degree- and weight-preserving
one-to-one correspondence
\begin{equation}\label{isd}
\begin{array}{ccc}
\mathcal I(P_M^p[J(\boldsymbol c)])
&
\overset{1:1}{\longleftrightarrow}
&
\{\st\in\cc(G)\mid \wt_{\boldsymbol c}(\st)\ne0\}
\\[0.2ex]
\rotatebox[origin=c]{90}{$\in$}
&&
\rotatebox[origin=c]{90}{$\in$}
\\[-0.4ex]
I=\{i_1,\ldots,i_k\}
&
\longleftrightarrow
&
\st_I,
\end{array}
\end{equation}
namely,
\begin{equation}\label{stbi}
|I|=\dd(\st_I),
\qquad
\wt_{\boldsymbol c}(\st_I)
=\hf_{i_1}\cdots\hf_{i_k}.
\end{equation}
Consequently,
\begin{equation}\label{tgex}
T_G(y;\boldsymbol c)
=
\sum_{I\in\mathcal I(P_M^p[J(\boldsymbol c)])}
y^{|I|}
\prod_{i\in I}\hf_i.
\end{equation}
Here the product over $i\in I$ is independent of the order, since
the corresponding elementary path weights commute by
Lemma~\ref{le:halg}.
\end{prop}

\begin{proof}
We first assume that all $c_i$ are nonzero, so that
$J(\boldsymbol c)=\{1,\ldots,M\}$ and hence
$P_M^p[J(\boldsymbol c)]=P_M^p$.
The boundary conditions and the nonzero local configurations in
Figure~\ref{f:los} imply that every state of nonzero weight is uniquely
decomposed into elementary paths of the two types in \eqref{hs}.
Each occupied vertical boundary edge belongs to exactly one such path.
It either passes straight through a column, giving an $A_j$,
or turns to the neighboring column along the bottom row, returns at
some row $a$, and then proceeds vertically upward, giving a $B_{j,a}$.
Two elementary paths are incompatible, that is, cannot coexist in a
state of nonzero weight, precisely in the following cases:
\begin{equation}
A_j\ \text{and}\ B_{j,a},\quad
B_{j,a}\ \text{and}\ A_{j+1},\quad
B_{j,a}\ \text{and}\ B_{j,b}\ (a\ne b),\quad
B_{j,a}\ \text{and}\ B_{j+1,b}\ (a>b).
\end{equation}
All other pairs are compatible.
Since the conservation law \eqref{wp} and the nonvanishing of the local
weights are both local conditions, a collection of elementary paths
occurs in a state of nonzero weight if and only if every pair in the
collection is compatible.
In terms of the ordering in \eqref{hs}, two elementary paths
$\hf_i$ and $\hf_j$ with $i<j$ are compatible precisely when
$i\prec j$ as defined in \eqref{prec}.
Thus the sets of indices arising from states of nonzero weight are
precisely the independent sets of $P_M^p$.

We next compare the weights.
Compatible elementary paths need not be disjoint.
The two local possibilities that require a check are illustrated in
Figure~\ref{fig:ov}.


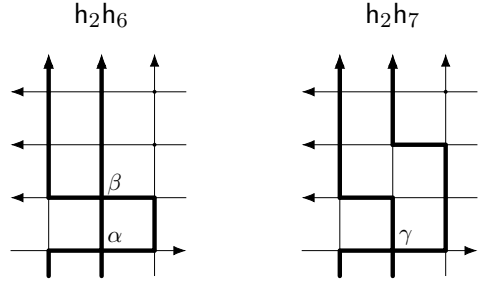
\begin{figure}[H]
\centering

\begin{tikzpicture}[
  scale=0.7,
  >={Latex[length=1.7mm,width=1.25mm]}
]

\begin{scope}[
  xshift=0cm,
  thinwire/.style={line width=.34pt},
  pathwire/.style={
    line width=1.65pt,
    line cap=round,
    line join=round
  }
]

\node at (2,4.45) {$\hf_2\hf_6$};

\foreach \x in {1,2,3}
  \draw[thinwire,->] (\x,-.48)--(\x,3.72);

\draw[thinwire,->] (.28,0)--(3.60,0);

\foreach \y in {1,2,3}
  \draw[thinwire,<-] (.28,\y)--(3.60,\y);

\foreach \x in {1,2,3}
  \foreach \y in {0,1,2,3}
    \fill (\x,\y) circle (1.05pt);

\node[scale=0.8] at (2.25,0.25) {$\alpha$};
\node[scale=0.8] at (2.25,1.25) {$\beta$};

\draw[pathwire]
  (1,-.48)--(1,0)--(2,0)--(2,1)--(1,1);

\draw[pathwire,-{Latex[length=2.0mm,width=1.45mm]}]
  (1,1)--(1,3.72);

\draw[pathwire]
  (2,-.48)--(2,0)--(3,0)--(3,1)--(2,1);

\draw[pathwire,-{Latex[length=2.0mm,width=1.45mm]}]
  (2,1)--(2,3.72);

\end{scope}

\begin{scope}[
  xshift=5.5cm,
  thinwire/.style={line width=.34pt},
  pathwire/.style={
    line width=1.65pt,
    line cap=round,
    line join=round
  }
]

\node at (2,4.45) {$\hf_2\hf_7$};

\foreach \x in {1,2,3}
  \draw[thinwire,->] (\x,-.48)--(\x,3.72);

\draw[thinwire,->] (.28,0)--(3.60,0);

\foreach \y in {1,2,3}
  \draw[thinwire,<-] (.28,\y)--(3.60,\y);

\foreach \x in {1,2,3}
  \foreach \y in {0,1,2,3}
    \fill (\x,\y) circle (1.05pt);

\node[scale=0.8] at (2.25,0.25) {$\gamma$};

\draw[pathwire]
  (1,-.48)--(1,0)--(2,0)--(2,1)--(1,1);

\draw[pathwire,-{Latex[length=2.0mm,width=1.45mm]}]
  (1,1)--(1,3.72);

\draw[pathwire]
  (2,-.48)--(2,0)--(3,0)--(3,2)--(2,2);

\draw[pathwire,-{Latex[length=2.0mm,width=1.45mm]}]
  (2,2)--(2,3.72);

\end{scope}

\end{tikzpicture}

\caption{Two examples of compatible elementary paths from
Figure~\ref{fig:p9}.}
\label{fig:ov}
\end{figure}

At $\alpha$, the factors contributed by $\hf_2$ and $\hf_6$ are
$W_\alpha$ and $W_\alpha^{-1}$, whose product is $1$.
This agrees with the four-occupied-edge weight in the lower row of
Figure~\ref{f:los}.
The same cancellation occurs at $\gamma$.
At $\beta$, the factors contributed by $\hf_2$ and $\hf_6$ are
$s_\beta W_\beta$ and $W_\beta^{-1}$, whose product is $s_\beta$.
This reproduces the corresponding four-occupied-edge weight in the
upper row of Figure~\ref{f:los}.

These are the only possibilities when compatible elementary paths
meet at a vertex.
For a pairwise compatible collection, the same local calculation
applies independently at every shared vertex.
Hence
$\wt_{\boldsymbol c}(\st_I)
=\hf_{i_1}\cdots\hf_{i_k}$.
Since each elementary path contains exactly one occupied vertical
boundary edge, the correspondence preserves the degree:
$\dd(\st_I)=|I|$.
This proves \eqref{stbi}, and summing over all states gives
\eqref{tgex} when all $c_i$ are nonzero.

For a general $\boldsymbol c$, a state has nonzero weight precisely
when none of its elementary paths has an index outside
$J(\boldsymbol c)$.
Indeed, if $i\notin J(\boldsymbol c)$, then $c_i=0$, so every state
containing the elementary path $\hf_i$ has zero weight.
Conversely, if all elementary paths occurring in a state have indices
in $J(\boldsymbol c)$, then the corresponding parameters $c_i$ are
nonzero, and the weight calculation above remains valid.
Thus the states of nonzero weight are indexed by
$\mathcal I(P_M^p[J(\boldsymbol c)])$.
The degree- and weight-preserving correspondence remains unchanged,
and summing over these states gives \eqref{tgex}.
\end{proof}

As an example of \eqref{tgex}, consider the case $p=L=3$
shown in Figure~\ref{fig:p9}. Then $M=9$.
For generic $\boldsymbol c$, so that
$J(\boldsymbol c)=\{1,\ldots,9\}$, we have
\begin{equation}\label{tex9}
T_G(y;\boldsymbol c)
=
1+y\sum_{i=1}^9\hf_i
+y^2\sum_{1\le i\prec j\le9}\hf_i\hf_j
+y^3\hf_1\hf_5\hf_9,
\end{equation}
where $i\prec j$ means $i+3<j$ according to \eqref{prec}.
The cubic term corresponds to the unique independent set of
cardinality three in $P_9^3$, namely $\{1,5,9\}$.

Now keep the same wiring diagram $G$ and specialize
$\boldsymbol c$ by setting $c_5=0$, while all other $c_i$ remain
nonzero.
Then $J(\boldsymbol c)=\{1,\ldots,9\}\setminus\{5\}$, and
\eqref{tgex} gives
\begin{equation}\label{tex8}
T_G(y;\boldsymbol c)
=
1+y\sum_{i\in J(\boldsymbol c)}\hf_i
+y^2
\sum_{\substack{i,j\in J(\boldsymbol c)\\ i\prec j}}
\hf_i\hf_j.
\end{equation}
Thus the terms indexed by independent sets containing $5$ disappear,
and the cubic term is absent.
Equivalently, the nonzero contributions are governed by the induced
subgraph $P_9^3[J(\boldsymbol c)]$ obtained by deleting the vertex $5$
from $P_9^3$.

\section{Q6V realization of indifference graphs and independent-set operators}
\label{s:main}

In this section, we specialize the Q6V model on $G_{p,L}$
introduced in the previous section to an odd root of unity.
The resulting elementary path weights satisfy the relations associated
with the path power $P_M^p$.
Combining this with Proposition~\ref{prp}, we realize arbitrary finite
indifference graphs within the Q6V model and identify the corresponding
independent-set operators with Q6V transfer matrices.

\subsection{Root-of-unity specialization}\label{ss:ru}

Let $N\ge3$ be odd, let $\ve$ be a primitive $N$th root of unity,
and put $\omega=\ve^{-2}$.  Then $\omega$ is also a primitive
$N$th root of unity.

Following \cite{IKTY26}, we specialize $q=\ve$ and realize each
$q$-Weyl algebra generated by $U_{j,a},W_{j,a}$ on an $N$-dimensional
cyclic representation satisfying
\begin{equation}\label{rep}
U_{j,a}^N=W_{j,a}^N=\mathrm{Id},\qquad
U_{j,a}W_{j,a}=\ve W_{j,a}U_{j,a}.
\end{equation}
A standard realization is
\begin{equation}\label{mat}
U_{j,a}=
\begin{pmatrix}
1&&&&\\
&\ve&&&\\
&&\ve^2&&\\
&&&\ddots&\\
&&&&\ve^{N-1}
\end{pmatrix}_{j,a},
\qquad
W_{j,a}=
\begin{pmatrix}
0&0&\cdots&0&1\\
1&0&\cdots&0&0\\
0&1&\ddots&0&0\\
\vdots&&\ddots&\ddots&\vdots\\
0&0&\cdots&1&0
\end{pmatrix}_{j,a}.
\end{equation}
More generally, one may use a representation obtained from this one
by pullback under an automorphism of the $\ve$-Weyl algebra preserving
\eqref{rep}.

Following the notation of \cite{IKTY26}, let
\begin{equation}\label{bbT}
\mathbb T_G(y;\boldsymbol c)
=
T_G(y;\boldsymbol c)|_{q=\ve}
\in\mathrm{End}((\C^N)^{\otimes K})
\end{equation}
denote the resulting specialization of the Q6V transfer matrix.
The commutativity of the Q6V transfer matrices survives this
specialization, so that
\begin{equation}\label{bbcom}
[\mathbb T_G(y;\boldsymbol c),
 \mathbb T_G(z;\boldsymbol c)]
=0.
\end{equation}

We continue to use the symbol $\hf_i$ for the elementary path weights.
From \eqref{rep}, \eqref{hs}, and \eqref{hcu},
\begin{equation}\label{pow}
\hf_i^N=c_i^N\mathrm{Id}\qquad(1\le i\le M).
\end{equation}
Moreover, Lemma~\ref{le:halg} becomes
\begin{equation}\label{rhal}
\hf_i\hf_j=
\begin{cases}
\omega\hf_j\hf_i,&i<j\le i+p,\\
\hf_j\hf_i,&i+p<j.
\end{cases}
\end{equation}
Thus \eqref{pow} and \eqref{rhal} have exactly the form of
\eqref{halg} associated with the path power $P_M^p$.
In particular, when all $c_i$ are nonzero, the frustration graph of
$\hf_1,\ldots,\hf_M$ is $P_M^p$.

\subsection{Q6V realization}\label{ss:mt}

We are now ready to state the main result of the paper.

\begin{thm}\label{th:main}
Let $\gf$ be a nonempty finite indifference graph with an indifference ordering
$v_1<\cdots<v_n$, and write $h_i$ for the operator associated with $v_i$.
Let $N\ge3$ be odd.
For nonzero scalars $b_1,\ldots,b_n$, the system \eqref{halg}
associated with $\gf$ admits a realization by the root-of-unity Q6V
model on some $G_{p,L}$.

More precisely, after reordering the connected components if necessary,
let $v_i\mapsto m_i$ be the realization of $\gf$ as
$P_M^p[\{m_1,\ldots,m_n\}]$ given by Proposition~\ref{prp}.
The remaining parameters in \eqref{spep} can then be further specialized so that:

\begin{enumerate}
\item
under the identification $h_i=\hf_{m_i}$, the relations
\eqref{halg} hold and
$\hf_{m_i}^N=b_i^N\mathrm{Id}$;

\item
the Q6V states of nonzero weight are in one-to-one correspondence with
$\mathcal I(\gf)$, with degree and weight preserved as in
\eqref{stbi};

\item
the independent-set operator \eqref{tran} and the transfer matrix
\eqref{bbT} are identified as
\begin{equation}\label{tid}
\mathcal T_{\gf}(y)
=
\mathbb T_G(y;\boldsymbol c).
\end{equation}
\end{enumerate}
\end{thm}

\begin{proof}
With the notation of the theorem, choose $L$ so that
$M=(p+1)(L-1)+1\ge m_n$.
We specialize the parameters $c_1,\ldots,c_M$ in \eqref{cdef} by
\begin{equation}\label{cb}
c_{m_i}=b_i\;\;(1\le i\le n),
\qquad
c_r=0\;\;
\bigl(r\notin\{m_1,\ldots,m_n\}\bigr),
\end{equation}
so that, by \eqref{jc},   
\begin{equation}\label{jcm}
J(\boldsymbol c)=\{m_1,\ldots,m_n\}.
\end{equation}
By \eqref{rhal}, the operators $\hf_{m_i}$ and $\hf_{m_j}$
$\omega$-commute precisely when $\{v_i,v_j\}$ is an edge of $\gf$,
and otherwise commute.
Equation~\eqref{pow} gives
$\hf_{m_i}^N=b_i^N\mathrm{Id}$.
Hence the identification $h_i=\hf_{m_i}$ satisfies \eqref{halg},
which proves (1).

By Proposition~\ref{pr:Tg}, the Q6V states of nonzero weight are in one-to-one
correspondence with $\mathcal I(P_M^p[J(\boldsymbol c)])$.
Under the graph isomorphism $v_i\mapsto m_i$,
$\mathcal I(P_M^p[J(\boldsymbol c)])$ is naturally identified with
$\mathcal I(\gf)$.
The claimed preservation of degree and weight follows from \eqref{stbi}.
This proves (2).

Finally, by (2) and the definitions \eqref{bbT} and \eqref{tran},
\begin{equation}
\mathbb T_G(y;\boldsymbol c)
=
\sum_{I\in\mathcal I(\gf)}
y^{|I|}h_I
=
\mathcal T_{\gf}(y).
\end{equation}
This proves \eqref{tid}, and hence (3).
\end{proof}

As an illustration of Theorem~\ref{th:main}, consider again the graph
in Figure~\ref{fex3}. We may take
$p=8$ and $L=3$, so that $M=19$, and use the embedding
\begin{equation}
\gf\simeq P_{14}^8[\{1,5,7,14\}]
\subset P_{19}^8.
\end{equation}
Setting
$
c_1=b_1,\,
c_5=b_2,\,
c_7=b_3,\,
c_{14}=b_4,
$
and all other $c_i$ equal to zero gives
$J(\boldsymbol c)=\{1,5,7,14\}$.
Hence Theorem~\ref{th:main} realizes the four generators
$h_1,h_2,h_3,h_4$ by
$\hf_1,\hf_5,\hf_7,\hf_{14}$, respectively.
In this case,
\begin{equation}
\mathbb T_G(y;\boldsymbol c)
=
\mathrm{Id}
+y(\hf_1+\hf_5+\hf_7+\hf_{14})
+y^2(\hf_1\hf_{14}+\hf_5\hf_{14}).
\end{equation}

\begin{cor}\label{co:fr}
For odd $N$, under the realization in Theorem~\ref{th:main}, the
commutativity relation \eqref{tcom} and the Frobenius relation
\eqref{frob} follow respectively from the Q6V identities
\cite[(3.10)]{IKTY26} and \cite[(3.14)]{IKTY26}.
\end{cor}

\begin{proof}
The commutativity relation \eqref{tcom} follows immediately from
\eqref{tid} and \eqref{bbcom}.
The root-of-unity Frobenius identity of \cite[(3.14)]{IKTY26},
applied to the parameter specialization in Theorem~\ref{th:main},
gives
\begin{equation}
\mathbb T_G(y;\boldsymbol c)
\mathbb T_G(\omega y;\boldsymbol c)
\cdots
\mathbb T_G(\omega^{N-1}y;\boldsymbol c)
=
\mathscr P_{G,N}(y^N)\,\mathrm{Id}.
\end{equation}
Here $\omega=\ve^{-2}$; since $N$ is odd, the $\omega$-orbit is the
same as the $\ve$-orbit used in \cite[(3.14)]{IKTY26}, and the ordering is
irrelevant by commutativity \eqref{bbcom}.
By the definition of $\mathscr P_{G,N}$ in
\cite[(3.9), (3.13)]{IKTY26} and of $\mathcal Z_{\gf,N}$ 
in \eqref{zgn},
the specialization in Theorem~\ref{th:main} gives
$\mathscr P_{G,N}(y^N)=\mathcal Z_{\gf,N}(y^N)$.
Combining this with \eqref{tid}, and noting that
$(-1)^{N+1}=1$ for odd $N$, we obtain \eqref{frob}.
\end{proof}

\begin{remark}[Non-homogeneous ranges]
The Q6V construction itself uses the fixed range $p$ of the path power
$P_M^p$.  After the selected vertices are relabeled consecutively,
however, Proposition~\ref{pr:nhr} in Appendix~\ref{app:nhr} 
shows that the resulting frustration
graph admits a non-homogeneous range representation satisfying
\eqref{nhrcond}.
Thus Theorem~\ref{th:main} also applies to the non-homogeneous
interaction-range realizations considered in~\cite{A25}.
\end{remark}

\section{Summary and discussion}\label{s:o}

We have shown that every finite indifference graph can be realized as
the frustration graph of a suitable Q6V model at odd roots of unity.
Under this realization, the corresponding independent-set operator is
identified with the Q6V transfer matrix, thereby providing a Q6V
realization of the graph-theoretic free-parafermion identities.
This result also clarifies the relation between the two frameworks
and their respective ranges of generality, as we discuss below.

For exchange relations of the form
$h_i h_j=\omega^a h_jh_i$ with $a$ restricted to $0,\pm1$,
the graph-theoretic framework of \cite{MEWC25} is more general in some
respects: it applies to arbitrary $N$ and is formulated
only in terms of algebraic relations among the operators $h_i$.
By contrast, the Q6V approach given here provides a specific
realization and is available only for odd $N$.
The graph-theoretic framework also provides explicit raising and lowering
operators associated with the free-parafermion spectrum.
How to construct their analogues for general Q6V models is not
known at present and remains an important problem.

On the other hand, the Q6V approach applies to a much broader class of
wiring diagrams than the particular family $G_{p,L}$ used in this paper.
In this more general setting, the exchange relations may involve
exponents $a\in\Z_N$ beyond $0,\pm1$.
To encode this additional information graph-theoretically, one would
have to decorate the oriented edges of the frustration graph by labels
in $\Z_N$ recording the corresponding powers of $\omega$.
In this sense, the Q6V model accommodates structures not captured by
the frustration graph alone.

\appendix

\section{An explicit rational embedding into a path power}
\label{app:embed}

For completeness, we give a direct construction of the embedding in
Proposition~\ref{prp}.
The construction is not needed for the existence statement of that
proposition, but is used in Section~\ref{ss:exp} to obtain an explicit
path-power embedding in the example considered there.

Let $\gf$ be a nonempty finite indifference graph with an indifference
ordering $v_1<\cdots<v_n$ in which every connected component forms a
consecutive block.
By Proposition~\ref{pr:ub}, applied componentwise, this ordering satisfies
the umbrella property \eqref{up}.

For each $i$, set
\begin{equation}\label{eli}
\ell_i=
\min\bigl(
\{i\}\cup
\{j<i\mid\{v_j,v_i\}\in E(\gf)\}
\bigr).
\end{equation}
Then
$N_i^-=\{v_{\ell_i},v_{\ell_i+1},\ldots,v_{i-1}\}$,
with the set empty when $\ell_i=i$.

Starting with $x_1=0$, define $x_i$ recursively for $2\le i\le n$ by
\begin{equation}\label{app:xrec}
x_i=
\begin{cases}
x_{i-1}+2,
&\ell_i=i,
\\[2pt]
\dfrac{
\max\{x_{i-1},x_{\ell_i-1}+1\}
+x_{\ell_i}+1}{2},
&1<\ell_i<i,
\\[6pt]
\dfrac{x_{i-1}+x_1+1}{2},
&\ell_i=1.
\end{cases}
\end{equation}
We claim that
\begin{subequations}\label{app:claim}
\begin{align}
&x_1<\cdots<x_n,
\label{app:claima}\\
&\begin{cases}
x_j-x_i<1,
&\{v_i,v_j\}\in E(\gf),\\
x_j-x_i>1,
&\{v_i,v_j\}\notin E(\gf),
\end{cases}
\qquad 1\le i<j\le n.
\label{app:claimb}
\end{align}
\end{subequations}

We prove \eqref{app:claima} and \eqref{app:claimb} simultaneously
by induction on $n$.
The case $n=1$ is immediate.
Assume that \eqref{app:claima} and \eqref{app:claimb} hold for
$v_1,\ldots,v_{n-1}$, and consider $x_n$ defined by \eqref{app:xrec}.

Suppose first that $\ell_n=n$.
Then $v_n$ has no left neighbors and $x_n=x_{n-1}+2$.
Hence $x_n>x_{n-1}$ and, for every $j<n$,
$x_n-x_j\ge x_n-x_{n-1}=2>1$.
Thus both \eqref{app:claima} and \eqref{app:claimb} hold.

Next suppose that $1<\ell_n<n$.
Since $v_{\ell_n}$ and $v_{n-1}$ both belong to the clique $N_n^-$,
they are adjacent or coincide.
Hence, by the induction hypothesis in the former case and trivially
in the latter,
$x_{n-1}<x_{\ell_n}+1$.
Also, by \eqref{app:claima} for $n-1$,
$x_{\ell_n-1}<x_{\ell_n}$.
Therefore \eqref{app:xrec} gives
\begin{equation}
\max\{x_{n-1},x_{\ell_n-1}+1\}
<
x_n
<
x_{\ell_n}+1.
\end{equation}
In particular, $x_n>x_{n-1}$.
Moreover, if $\ell_n\le j<n$, then
$x_n-x_j\le x_n-x_{\ell_n}<1$,
whereas if $j<\ell_n$, then
$x_n-x_j\ge x_n-x_{\ell_n-1}>1$.
By the definition of $\ell_n$, the vertices
$v_{\ell_n},\ldots,v_{n-1}$ are precisely the left neighbors of
$v_n$, while $v_1,\ldots,v_{\ell_n-1}$ are the left nonneighbors.
Thus \eqref{app:claimb} also holds for all pairs involving $v_n$.

Finally, suppose that $\ell_n=1$.
Then every $v_j$ with $j<n$ is adjacent to $v_n$.
Since $v_1$ and $v_{n-1}$ are adjacent or coincide, the induction
hypothesis in the former case, and the trivial inequality in the latter,
give $x_{n-1}<x_1+1$.
Hence \eqref{app:xrec} gives
$x_{n-1}<x_n<x_1+1$.
Thus $x_n>x_{n-1}$ and, for every $j<n$,
$x_n-x_j\le x_n-x_1<1$.
This completes the induction and proves
\eqref{app:claima} and \eqref{app:claimb}.

The choices in \eqref{app:xrec} are not unique.
For example, in the case $\ell_i=i$, any rational value satisfying
$x_i>x_{i-1}+1$ would serve in place of $x_{i-1}+2$.
Likewise, in the other two cases, the midpoint may be replaced by any
rational point lying strictly between the corresponding lower and
upper bounds.
The particular choices above are made only for definiteness.

All $x_i$ are rational.
Write $x_i=a_i/b_i$ in reduced form, with $b_i\ge1$, and set
\begin{equation}
D=\operatorname{lcm}(b_1,\ldots,b_n),
\qquad
p=D,
\qquad
m_i=Dx_i+1.
\end{equation}
By \eqref{app:claima},
$1=m_1<\cdots<m_n$, and
$m_j-m_i=D(x_j-x_i)$.
For an edge, \eqref{app:claimb} and the integrality of $m_j-m_i$ give
$1\le m_j-m_i\le D-1\le p$,
whereas for a nonedge,
$m_j-m_i>D=p$.
Therefore
\begin{equation}
\{v_i,v_j\}\in E(\gf)
\quad\Longleftrightarrow\quad
1\le m_j-m_i\le p
\qquad (1\le i<j\le n),
\end{equation}
which gives the explicit induced path-power embedding of
Proposition~\ref{prp}.

\section{Non-homogeneous range representations}\label{app:nhr}

We relate indifference graphs to the non-homogeneous interaction ranges
considered in~\cite{A25}.
Let $v_1<\cdots<v_n$ be an ordering of the vertices of a finite graph.
A \emph{non-homogeneous range representation} is specified by
nonnegative integers $p_1,\ldots,p_n$ such that
\begin{equation}\label{nhredge}
\{v_i,v_j\}\in E(\gf)
\quad\Longleftrightarrow\quad
0<j-i\le p_i
\qquad (i<j).
\end{equation}
We impose 
\begin{equation}\label{nhrcond}
p_{i+1}\ge p_i-1
\qquad (1\le i<n),
\end{equation}
which is the condition considered in~\cite{A25}.

\begin{prop}[Non-homogeneous ranges and indifference graphs]\label{pr:nhr}
A nonempty finite graph is an indifference graph if and only if it
admits a non-homogeneous range representation satisfying \eqref{nhrcond}.
\end{prop}

\begin{proof}
Suppose first that $\gf$ is an indifference graph.
Choose an indifference ordering and place the connected components
consecutively, preserving the order within each component.
Then take the embedding
$1=m_1<\cdots<m_n$ from Proposition~\ref{prp}.
Set
\begin{equation}
R_i=\max\{j\mid m_j-m_i\le p\},
\qquad
p_i=R_i-i.
\end{equation}
Since the thresholds $m_i+p$ increase, $R_i$ is nondecreasing.
By \eqref{emb}, adjacency for $i<j$ is equivalent to $j\le R_i$,
and hence to $0<j-i\le p_i$, which gives \eqref{nhredge}.
Moreover,
\begin{equation}
p_{i+1}-p_i=R_{i+1}-R_i-1\ge-1,
\end{equation}
so \eqref{nhrcond} holds.
Conversely, suppose that \eqref{nhredge} and \eqref{nhrcond} hold.
Set
\begin{equation}
R_i=\min\{n,i+p_i\}.
\end{equation}
Then $R_1\le\cdots\le R_n$, and adjacency for $i<j$ is equivalent to
$j\le R_i$.
Now let $i<j<k$ and suppose that $\{v_i,v_k\}\in E(\gf)$.
Then $k\le R_i\le R_j$.
Since $i<j<k$, the inequality $j\le R_i$ implies
$\{v_i,v_j\}\in E(\gf)$, while $k\le R_j$ implies
$\{v_j,v_k\}\in E(\gf)$.
Thus the ordering $v_1<\cdots<v_n$ satisfies the umbrella property
\eqref{up}. Hence $\gf$ is an indifference graph by
Proposition~\ref{pr:ub}, applied to each connected component.
\end{proof}

Thus, for finite graphs, the following three descriptions are equivalent:
indifference graphs, induced subgraphs of path powers, and
non-homogeneous range representations satisfying \eqref{nhrcond}.

\section*{Acknowledgments}
AK is deeply grateful to the late Rodney James Baxter for his generous
support and kindness over many years, both professionally and personally.
He also thanks Rei Inoue, Yuji Terashima, and Junya Yagi for their collaboration
on related earlier works.
AK is supported by JSPS KAKENHI Grant Number 24K06882.
Generative AI tools were used on a limited basis for language editing,
literature searches, and preliminary mathematical discussions.
All mathematical content and references were independently verified by the authors.


\end{document}